\documentclass[doublespacing]{elsart}

 \usepackage{graphics}

\usepackage{amssymb}

\begin{document}

\begin{frontmatter}


\title{Microwave Diagnostics of  Ultracold Neutral Plasma}


\author{Lu Ronghua, Guo Li, Han shensheng}

\address{Key
Laboratory for Quantum Optics and Center for Cold Atom Physics,
Shanghai Institute of Optics and Fine, Mechanics, Chinese Academy of
Sciences, Shanghai 201800, China}
\begin{abstract}
The microwave method is suggested to diagnose the ultracold neutral
plasma. Based on the calculations of the dipole radiation , we
derived the microwave scattering cross section of the ultracold
neutral plasma. The significant results indicate that we can
diagnose the total electron number and recombination rate by this
method.
\end{abstract}

\end{frontmatter}


\section{INTRODUCTION}
The ultracold neutral plasma (UNP)  is first generated by
photoionizing a ultracold gas\cite{Killian1999PRL}, in which the
typical electron and ion temperature are around $1\sim1000K$ and
$1K$ respectively. UNP  extends greatly the boundaries of classical
neutral plasma physics and has been widely studied in recent
years\cite{Killian2007PR}\cite{Killian2006SCI}. The diagnostics
methods of UNP are mainly developed from  some well defined
technique of  optic probes, such as laser induced fluorescence
imaging\cite{Cumings2005PRL}, optical absorption
imaging\cite{Simien2004PRL} and even recombination
fluorescence\cite{Bergeson2008PRL}. The optic probes can excite the
fluorescence or be absorbed  in the case of ions and Rydberg atoms.
However, free electrons in UNP can not respond to the laser beam
except the Thomason scattering which is too weak to be measured in
the current techniques. Furthermore, most traditional plasma
diagnostics  are not accessible due the size-limited UNP. In this
paper, we suggest a new method of using microwave radiation for the
study of UNP. Using this method, we can measure the amount of
electrons $N_e$  and the recombination rate of plasma, which is
extremely important in the research of UNP.

\section{DISCUSSION}
\begin{figure}
\begin{center}
\scalebox{0.5}{\includegraphics{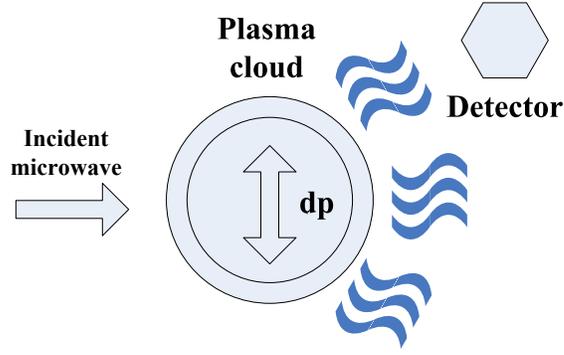}}
\caption{\label{fig:scattering} The physical scheme of the dipole radiation of UNP in an
incident microwave field }
\end{center}
\end{figure}

Normally, when the microwave wavelength $\lambda$ is much smaller
than the size plasma size $L$, microwave-based diagnostics such as
interferometry and reflectometry have found very broad application
in classical plasma diagnostics. However, the UNP size limited by
the beamwidth of the cooling laser is usually very small around the
range of $mm$ . In the small plasma objects situation
$\lambda/L\gg1$, those microwave diagnostics based on propagation,
absorption or reflection fail to work. Shneider and Miles develop
the theory in the case of small plasma objects by measuring the
radiation scattered by the effective oscillating plasma
dipole\cite{Shneider2005JAP}\cite{Zhang2006JAP}. In their work
\cite{Shneider2005JAP}, the plasma is static uniform plasma column,
and the microwave frequency  used is smaller than plasma frequency
$\omega<\omega_{pe}$. However, UNP is an expanding spherical plasma
cloud with radial inhomogeneity, the density profile of UNP decay
radically quickly in the few of $mm$. There is still no any
discussion about microwave diagnostics on this special plasma so
far. In this paper we investigate the dipole radiation of UNP in an
incident microwave as the physical scheme shown in
Fig.\ref{fig:scattering}.
%

From linearized electron motion equation, we can easily derive the
high frequency conductivity $\sigma$ and dielectric constant
$\epsilon$ of plasma\cite{WilliamBook}:
\begin{eqnarray}
\sigma &=& \frac{i\omega_{pe}^2}{4\pi \omega} \label{eps2}\\
\epsilon &=&
1+\frac{i4\pi\sigma}{\omega}=1-\frac{\omega_{pe}^2}{\omega^2}
\label{Eq:eps}
\end{eqnarray}
where $\omega_{pe}=\sqrt{n_ee^2/m_e\epsilon_0}$ is the plasma
frequency,$m_e$ is the electron mass and $n_e$ is eletron density .

We calculate the dipole of an uniform dielectric ball with the
radius $r$ and dielectric constant epsilon $\epsilon$  in the first
step. The dielectric ball responds to the incident electric field
$\vec{E}$ and the electric dipole is induced . we can get that
outside the sphere the potential is equivalent to that of the
applied field plus the field of a point electric dipole
$\vec{p}_{uniform}$, so the equivalent dipole of the dielectric ball
is
\begin{equation}
\vec{p}_{uniform}=\frac{\epsilon-1}{\epsilon+2} 4\pi\epsilon_0 r^3 \vec{E}
\label{Eq:uniformball}
\end{equation}

Next, we consider a thin spherical shell with spherical shell of
radius $r$, thickness $dr$ and dielectric constant $\epsilon(r)$,
From Eq.\ref{Eq:uniformball}, the corresponding dipole of the shell
is
\begin{equation}
d\vec{p}(r)=\frac{\epsilon(r)-1}{\epsilon(r)+2} 12\pi\epsilon_0  r^2 dr
\vec{E}
\label{Eq:thinshell}
\end{equation}

So the integral on the Eq.\ref{Eq:thinshell} along the radius yields
the equivalent dipole  of the UNP ball
\begin{eqnarray}
\vec{p}=\int d\vec{p}(r)=
\int \frac{
-\frac{\omega_{pe}(r)^2}{\omega^2}}{3-\frac{\omega_{pe}(r)^2}{\omega^2}
} 12\pi\epsilon_0 r^2 dr \vec{E}
\end{eqnarray}

Because the size of UNP is much less than the wavelength of
microwave, we can neglect the field variance across the UNP ball. We
set $\vec{E}=E_0  \vec{e_r} \exp{i\omega t}$ , the equivalent dipole
of the UNP ball will vibrate in the same frequency $\omega$ of the
incident electric field. The oscillation  mainly comes from the
oscillation of $\vec{E}$, so we  get total  dipole radiation power
in space
\begin{equation}
P=\frac{|\vec{p''}| ^2}{12\pi\epsilon_0 c^3}= \frac{12\pi\epsilon_0
E_0^2}{c^3} [ \int \frac{
-\omega_{pe}^2(r)}{3-\frac{\omega_{pe}^2(r)}{\omega^2} } r^2 dr ]^2
\label{Eq:Pwithomega}
\end{equation}
and the effective scattering cross section
\begin{equation}
\sigma(\omega)=\frac{P}{0.5 \epsilon_0cE_0^2}=\frac{24\pi}{c^4} [
\int \frac{ -\omega_{pe}^2(r)}{3-\frac{\omega_{pe}^2(r)}{\omega^2} }
r^2 dr ]^2 \label{Eq:sigmawithomega}
\end{equation}
$\omega_{pe}^2(r)$ can be written as $\omega^2_{pe0} f_e(r)$, where
$\omega_{pe0}$ is plasma frequency at the center of UNP and $f_e(r)$
is the radial density profile.

\begin{figure}
\begin{center}
\scalebox{0.5}{\includegraphics{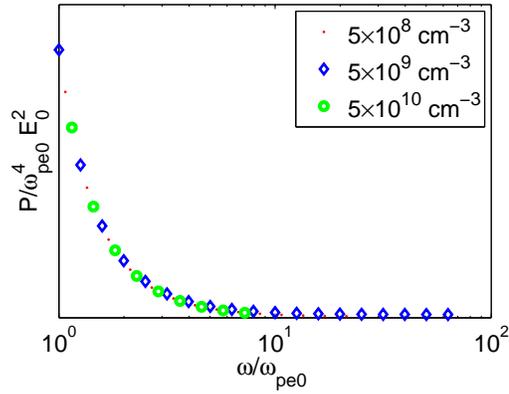}}
\caption{\label{fig:calRadiationPowerVsOmega2} Dipole radiation
power to $\omega_{pe0}^4E_0^2$ ratios at  different density $n_{e0}=5\times
10^8 cm^{-3},5\times 10^9 cm^{-3},5\times 10^{10} cm^{-3}$ }
\end{center}
\end{figure}
\begin{figure}
\begin{center}
\scalebox{0.5}{\includegraphics{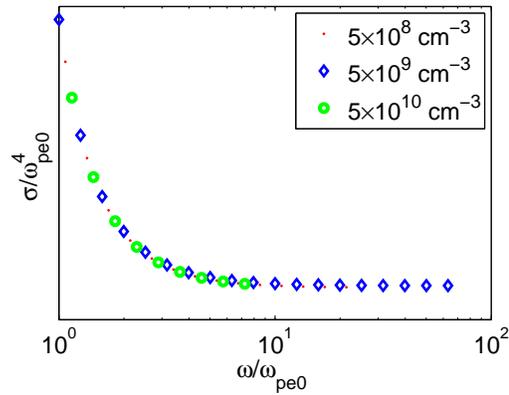}}
\caption{\label{fig:calCrossSectionVsOmega} Effective scattering
cross section to $\omega_{pe0}^4$ ratios at  different density
$n_{e0}=5\times 10^8 cm^{-3},5\times 10^9 cm^{-3},5\times 10^{10}
cm^{-3}$  }
\end{center}
\end{figure}

In Eq.\ref{Eq:Pwithomega} and Eq.\ref{Eq:sigmawithomega}, the dipole
radiation power and the effective scattering cross section tend to a
constant when the microwave frequency is much greater than the
plasma frequency. But it is worthy to note the corresponding
constant reflects the crucial properties of plasma. The UNP has a
typical guassian-like radial density profile
$f_e(r)=exp(-r^2/2r_0^2)$. In the UNP situation,
Fig.\ref{fig:calRadiationPowerVsOmega2} and
Fig.\ref{fig:calCrossSectionVsOmega} illuminate the constant
tendency when the ratio of microwave and plasma frequency is greater
than 5 in three  different density cases.  In the two figures, the
dipole radiation power $P(\omega)$ and the effective scattering
cross section $\sigma(\omega)$  are scaled vertically  by
$\omega_{pe0}^4E_0^2$ and $\omega_{pe0}^4$ respectively. The
overlapping horizontal lines at large frequency ratio indicate the
radiation power and cross section never change while $\omega$
increases at large $\omega$. The classical $\omega^4$ dependence of
dipole radiation is not satisfied here. Specially, when the
microwave frequency is large enough it shows the frequency
independence as we explained above. The underlying reason is the
frequency dependence of the equivalent dipole instead of the
assumption of constant dipole in the classical description.

For large $\omega/\omega_{pe0}$,  we can ignore the term of
${\omega_{pe}^2(r)}/{\omega^2}$ in Eq.\ref{Eq:Pwithomega} and
Eq.\ref{Eq:sigmawithomega}, and get the dipole radiation power
\begin{equation}
P= \frac{4\pi\epsilon_0 E_0^2}{3c^3} [ \int \omega_{pe}^2(r) r^2 dr
]^2= \frac{4\pi e^4 E_0^2}{3\epsilon_0 m_e^2c^3} [ \int n_e(r)  r^2
dr ]^2= \frac{ e^4 N_e^2E_0^2}{12\pi\epsilon_0 m_e^2c^3}
\label{Eq:Pwithoutomega}
\end{equation}
 the effective scattering cross section
\begin{equation}
\sigma=\frac{8\pi}{3c^4} [ \int \omega_{pe}^2(r)  r^2 dr
]^2=\frac{8\pi e^4}{3c^4m_e^2\epsilon_0^2}[ \int n_e(r)  r^2 dr
]^2=\frac{e^4 N_e^2}{6 \pi c^4m_e^2\epsilon_0^2}
\label{Eq:sigmawithoutomega}
\end{equation}
and the radiation electric field $E_r$ at distance $R$ and direction
$\phi$
\begin{equation}
E_r(R,\phi)=\frac{sin\phi}{2R}\sqrt{\frac{3P}{\pi\epsilon_0 c} }
=\frac{e^2N_e}{4\pi \epsilon_0c^2 R m_e}E_0\sin\phi
\label{Eq:Erwithoutomega}
\end{equation}
where $N_e=\int n_e(r) \ 4 \pi r^2 dr$ is the total number of
electron.

Eqns.\ref{Eq:Pwithoutomega}-\ref{Eq:Erwithoutomega}
 demonstrate the formula of UNP dipole radiation is free from the  the frequency $\omega$ or
plasma density profile $f_e(r)$, but only dependent of  total
electron number $N_e$. Though we still don't break the resolution
limitation on the microwave diagnostics technology, it is
significant to give a clue of the space integral on density i.e.
$N_e$.  Moreover, the recombination rate can be calculated based the
$N_e$ measurement.

The ultracold neutral plasma  is produced in  very low electron
temperature. In this range of temperature, the three body
recombination(TBR) dominates over the electron and ion recombination
. Through the TBR, one ion recombines with two electrons into a
Rydberg atom and an leftover electron  with the extra energy.  The
classical TBR theory predicts the recombination rate per ion is
$K_{TBR}\approx3.8\times10^{-9}T_e^{-9/2}n_e^2 \ s^{-1}$, where
$T_e$ is given in $K$ and the density $n_e$ in $cm^{-3}$ , so the
TBR rate varies with temperature as $T_e^{-9/2}$ and is very fast at
ultracold temperatures\cite{Killian2007PR}.  The TBR effect is very
important in UNP, because it is the main heating mechanism for the
electron at low $T_e$ \cite{Robicheaux2002PRL}. It has attracted
wide interest and  much controversy. Microwave diagnostics of  UNP
may offer a new way to detect the recombination rate of UNP.
%

\begin{figure}
\begin{center}
\scalebox{0.5}{\includegraphics{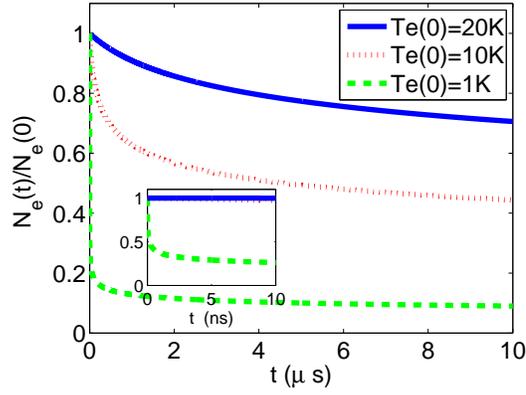}} \caption{\label{fig:Ne_t}
The evolution total electron number $N_e$ at different initial
$T_e$.  $N_e$ decrease more rapidly with lower $T_e$. }
\end{center}
\end{figure}

As one example of the application, we consider an expanding UNP with
$r_0=2 mm$ and initial center density $n_{e0}=5\times 10^9 cm^{-3}$.
These two parameters are determined by the cool laser and are
typical in the experiment. Though in general case the plasma cloud
expands with characteristic expansion time $\tau=\sqrt{m_i
r_0^2(0)/k_b[T_e(0)+T_i(0)]}$ after creation\cite{Laha2007PRL}, the
total electron number $N_e$ would not been  changed during the
expansion if there was no any recombination.   So when we calculate
the total electron number $N_e(t)$, we need  consider only TBR
without expansion. We use a simple formulations for  TBR
$K_{TBR}\approx3.8\times10^{-9}T_e^{-9/2}n_e^2 \ s^{-1}$ (There is
some other discussion about TBR \cite{Bergeson2008PRL}, but the
details of TBR mechanism are out of this paper's scope ). So we can
get $N_e(t)= N_e(0)\prod(1-K_{TBR}\Delta t)$, and
$T_e(t)=T_e(0)\prod(1+K_{TBR}\Delta t)$ . Fig.\ref{fig:Ne_t} shows
the typical time evolution of $N_e(t)$  at three distinct initial
electron temperatures. Clearly $N_e$  falls  too sharply in the
initial short time when  initial electron temperature $T_e(0)=1K$,
 it is hard to get enough temporal resolution.  So in the next
figure, we only plot curves in $T_e(0)=10K $ and $20K$ condition.

\begin{figure}
\begin{center}
\scalebox{0.5}{\includegraphics{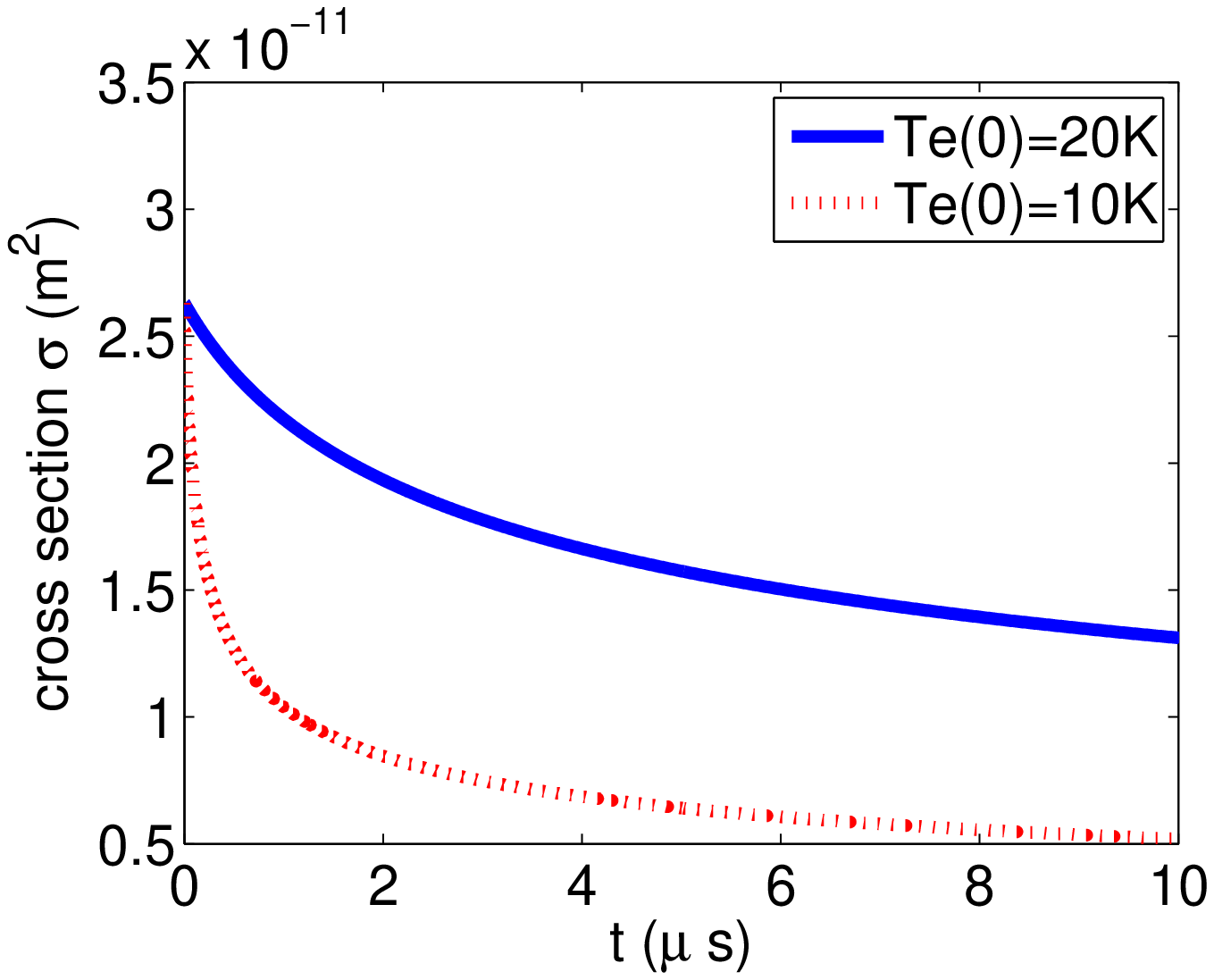}}
\scalebox{0.5}{\includegraphics{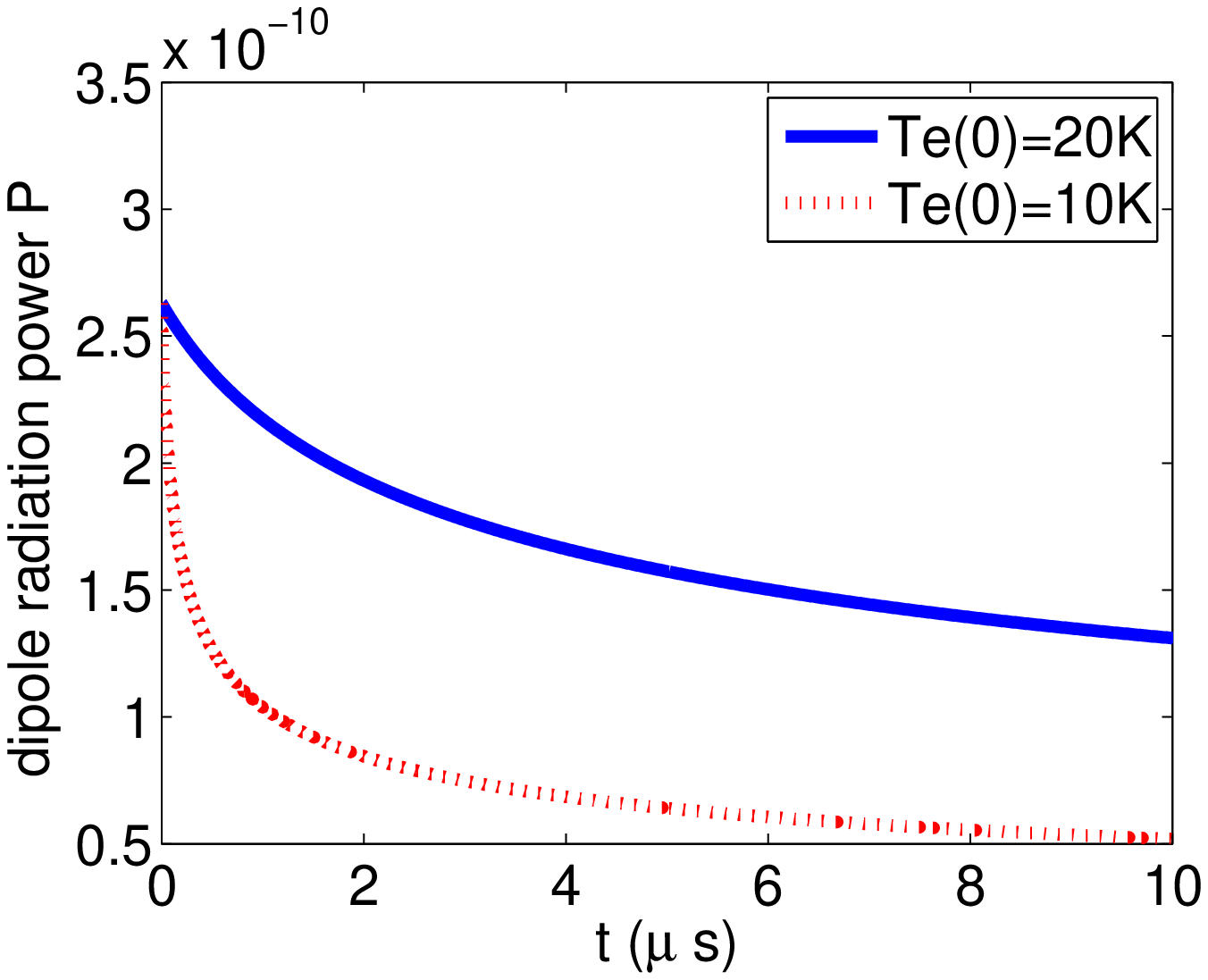}}
\scalebox{0.5}{\includegraphics{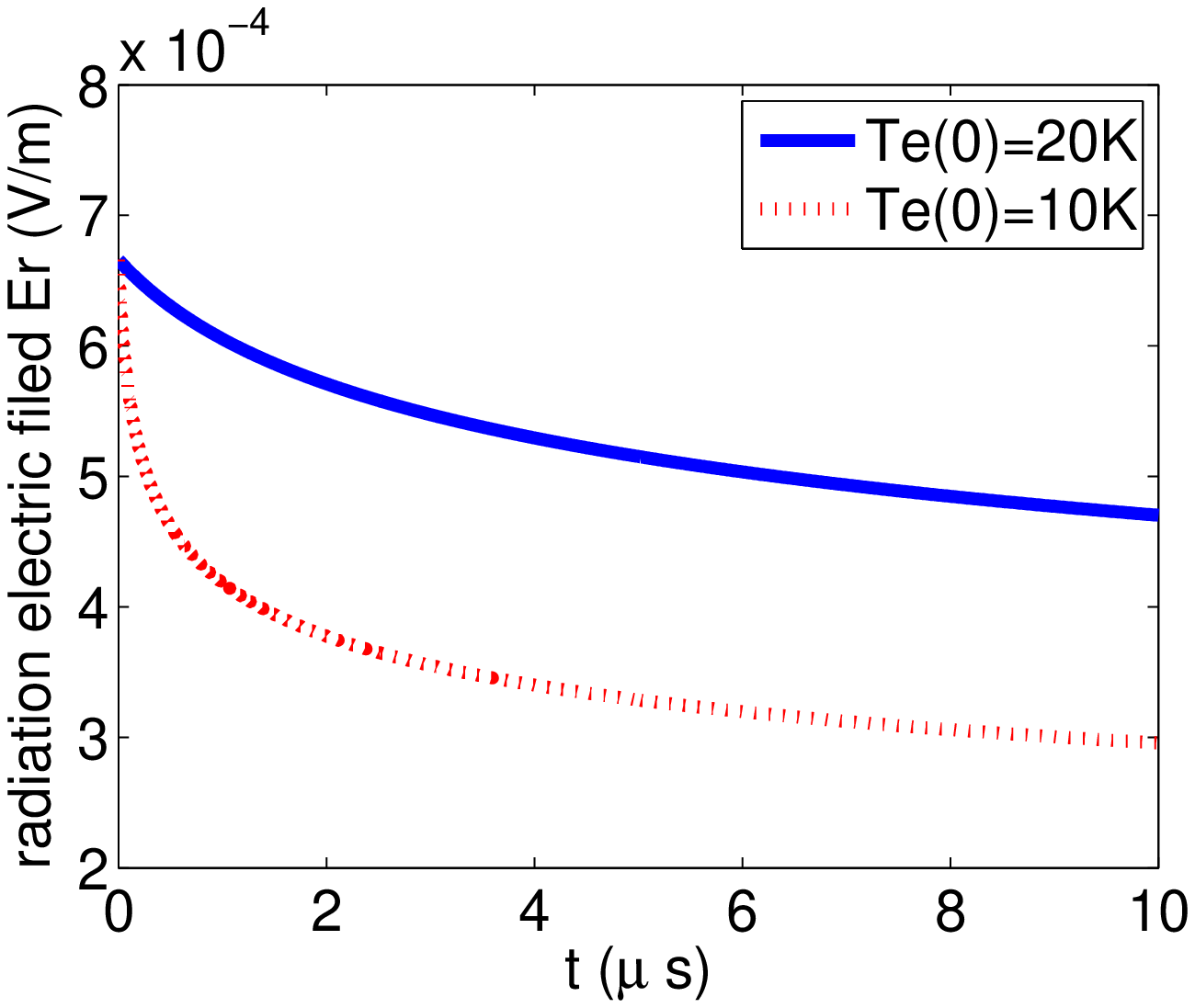}}
\caption{\label{fig:example} Effective cross section, radiation power and radiation
electric filed when  the intensity of the incident microwave equals
$10W/m^2$.}
\end{center}
\end{figure}

 Finally, we give a numerical case. The initial center density
 $n_{e0}=5\times 10^9 cm^{-3}$, so the plasma frequency at UNP
center $f_{pe0}\approx0.6GHz$. If incident microwave frequency $f$
is greater than $5f_{pe0}=3GHz$, we can directly use
Eqns.\ref{Eq:Pwithoutomega}-\ref{Eq:Erwithoutomega} to estimate the
equivalent dipole of the UNP cloud unrelated to $f$ .
Fig.\ref{fig:example} shows the results of effective cross section,
radiation power and radiation electric filed when the incident
microwave is $I_0=10W/m^2$, the radiation electric filed $E_r$ in
the fig is calculated at  $\phi=\pi/3$ and $r=0.2m$.

\section{SUMMARY}
Our calculations indicate that   the dipole radiations of UNP do not
depend on specific density profile $n_e(r)$ and incident frequency
$\omega$ when $\omega\gg\omega_{pe0}$ , but on the total electron
number $N_e$. We suggest that the microwave radiation from UNP may
offer a new  way to get the information of recombination in UNP, or
other  expanding inhomogeneous plasma  in similar.

\ack{The authors acknowledge the support of  National Natural
Science Foundation of China (Grant No.  10705042 ) }


\newpage
\emph{\textbf{{Figure captions}}}

Fig.1   The physical scheme of the dipole radiation of UNP in an
incident microwave field.

Fig.2   Dipole radiation power to $\omega_{pe0}^4E_0^2$ ratios at
different density $n_{e0}=5\times 10^8 cm^{-3},5\times 10^9
cm^{-3},5\times 10^{10} cm^{-3}$.

Fig.3 Effective scattering cross section to $\omega_{pe0}^4$ ratios
at  different density $n_{e0}=5\times 10^8 cm^{-3},5\times 10^9
cm^{-3},5\times 10^{10} cm^{-3}$.

Fig.4 The evolution total electron number $N_e$ at different initial
$T_e$.  $N_e$ decrease more rapidly with lower $T_e$.

Fig.5  Effective cross section, radiation power and radiation
electric filed when  the intensity of the incident microwave equals
$10W/m^2$.

\newpage


\end{document}